\documentclass[prb,twocolumn,showpacs]{revtex4}     
\usepackage{epsfig,amsmath} 

\begin{document}

\title{Thermoelectric Power of Dirac Fermions in Graphene} 

\author{Xin-Zhong Yan,$^{1}$ Yousef Romiah,$^2$ and C. S. Ting$^2$}
\affiliation{$^{1}$Institute of Physics, Chinese Academy of Sciences, P.O. Box 603, 
Beijing 100190, China\\
$^{2}$Texas Center for Superconductivity, University of Houston, Houston, Texas 77204, USA}
 
\date{\today}
 
\begin{abstract}
On the basis of self-consistent Born approximation for Dirac fermions under charged impurity scatterings in graphene, we study the thermoelectric power using the heat current-current correlation function. The advantage of the present approach is its ability to effectively treat the low doping case where the coherence process involving carriers in both upper and lower bands becomes important. We show that the low temperature behavior of the thermoelectric power as function of the carrier concentration and the temperature observed by the experiments can be successfully explained by our calculation.
\end{abstract}

\pacs{73.50.Lw, 73.50.-h, 72.10.Bg, 81.05.Uw} 

\maketitle

\section{Introduction}

Recent experimental observations \cite{Zuev,Wei,Ong} have revealed the unusual behavior of the thermoelectric power $S$ as function of the carrier concentration $n$ in graphene at low temperature. Near zero carrier concentration, the observed result of $S$ explicitly departures from the formula $\propto 1/\sqrt{n}$ given by the semiclassical Boltzmann theory \cite{Mott,Peres,Stauber} with $n$ as the number density of the charge carriers. Instead of diverging at $n = 0$, $S$ varies dramatically but continuously with changing sign as $n$ varying from hole side to electron side. There exist phenomenological explanations on this problem.\cite{Zuev,Hwang1} Though the quantum mechanical calculations based on the short-range impurity scatterings \cite{Lofwander,Dora} can qualitatively explain the behavior of the thermal-electric power, they cannot produce the linear-carrier-density dependence of the electric conductivity. Since the thermoelectric power is closely related to the electric transport, a satisfactory microscopic model dealing with the two problems in a self-consistent manner is needed. So far, such a microscopic theory for the thermal and electric transport of Dirac fermions in graphene is still lacking. 

It has well been established that the charged impurities in graphene are responsible for the carrier density dependences of the electric conductivity \cite{Nomura,Hwang,Yan} and the Hall coefficient \cite{Yan1} as measured in the experiments by Novoselov {\it et al.}.\cite{Geim} In the present work, based on the conserving approximation within the self-consistent Born approximation (SCBA), we develop the theory for the thermoelectric power $S$ of the Dirac fermions in graphene using the heat current-current correlation function under the scatterings due to charged impurities. This approach has been proven to be effective in treating the electric transport property of graphene at low carrier density,\cite{Yan,Yan1} there the coherence between the upper and lower bands is automatically taken into account. It is the coherence that yields finite minimum conductivity at zero carrier density. We will calculate the thermoelectric power as function of carrier concentration at low temperature and compare with the experimental measurements.\cite{Zuev,Wei,Ong} 

\section{Formalism}

We start with description of the electrons in graphene. At low carrier concentration, the low energy excitations of electrons in graphene can be viewed as massless Dirac fermions \cite{Wallace,Ando,Castro,McCann} as being confirmed by recent experiments.\cite{Geim,Zhang} Using the Pauli matrices $\sigma$'s and $\tau$'s to coordinate the electrons in the two sublattices ($a$ and $b$) of the honeycomb lattice and two valleys (1 and 2) in the first Brillouin zone, respectively, and suppressing the spin indices for briefness, the Hamiltonian of the system is given by
\begin{equation}
H = \sum_{k}\psi^{\dagger}_{k}v\vec
 k\cdot\vec\sigma\tau_z\psi_{k}+\frac{1}{V}\sum_{kq}\psi^{\dagger}_{k-q}V_i(q)\psi_{k} \label{H}
\end{equation}
where $\psi^{\dagger}_{k}=(c^{\dagger}_{ka1},c^{\dagger}_{kb1},c^{\dagger}_{kb2},c^{\dagger}_{ka2})$ is the fermion operator, $v$ ($\sim$ 5.86 eV\AA) is the velocity of electrons, $V$ is the volume of system, and $V_i(q)$ is the electron-impurity interaction. Here, the momentum $k$ is measured from the center of each valley with a cutoff $k_c \approx \pi/3a$ (with $a \sim 2.4$ \AA~the lattice constant), within which the electrons can be regarded as Dirac particles. By neglecting the intervalley scatterings that are unimportant here, $V_i(q)$ reduces to $n_i(-q)v_0(q)\tau_0\sigma_0$ with $n_i(-q)$ and $v_0(q)$ as respectively the Fourier components of the impurity density and the electron-impurity potential. For the charged impurity, $v_0(q)$ is given by the Thomas-Fermi (TF) type
\begin{equation}
v_0(q)=\frac{2\pi e^2}{(q+q_{TF})\epsilon}\exp(-qR_i)   \label{vi}
\end{equation}
where $q_{TF} = 4k_Fe^2/v\epsilon$ is the TF wavenumber, $k_F = \sqrt{\pi n}$ (with $n$ as the carrier density) is the Fermi wavenumber, $\epsilon \sim 3$ is the effective dielectric constant, and $R_i$ is the distance of the impurity from the graphene layer. This model has been successfully used to study the electric conductivity \cite{Yan} and the Hall coefficient.\cite{Yan1} As in the previous calculation, we here set $R_i = 0$ and the average impurity density as $n_i = 1.15\times 10^{-3}a^{-2}$.

\begin{figure} 
\centerline{\epsfig{file=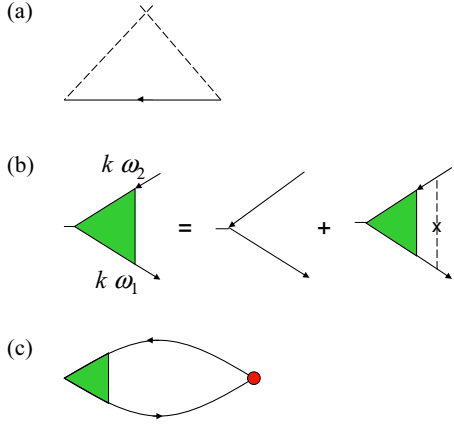,width=6.5 cm}}
\caption{(color online) (a) Self-consistent Born approximation for the self-energy of the single Dirac fermion. (b) Current vertex with impurity insertions. (c) Heat current-current correlation function. The red solid circle denotes the heat current vertex.} 
\end{figure} 

Under the SCBA [see Fig. 1(a)],\cite{Fradkin,Lee1} the Green function $G(k,\omega)=[\omega+\mu-v\vec
 k\cdot\vec\sigma\tau_z-\Sigma(k,\omega)]^{-1}$ $\equiv g_0(k,\omega)+ g_c(k,\omega)\hat k\cdot\vec\sigma\tau_z$ and the self-energy $\Sigma(k,\omega)=\Sigma_0(k,\omega)+\Sigma_c(k,\omega)\hat k\cdot\vec\sigma\tau_z$ of the single particles are determined by coupled integral equations:\cite{Yan}
\begin{eqnarray}
\Sigma_0(k,\omega) &=& \frac{n_i}{V}\sum_{k'}v^2_0(|k- k'|)g_0(k',\omega)\label{b1}\\
\Sigma_c(k,\omega) &=& \frac{n_i}{V}\sum_{k'}v^2_0(|k-k'|)
g_c(k',\omega)\hat k\cdot\hat k'    \label{b2}\\
g_0(k,\omega) &=& \frac{1}{2}[g_+(k,\omega)+g_-(k,\omega)] \label{b3}\\
g_c(k,\omega) &=& \frac{1}{2}[g_+(k,\omega)-g_-(k,\omega)] \label{b4}
\end{eqnarray}
where $g_{\pm}(k,\omega)=[\omega+\mu\mp vk-\Sigma_0(k,\omega)\mp\Sigma_c(k,\omega)]^{-1}$ with $\mu$ the chemical potential, $\hat k$ is the unit vector in $k$ direction, and the frequency $\omega$ is understood as a complex quantity. Here $g_{+}(k,\omega)$ [or $g_{-}(k,\omega)$] can be viewed as the upper (lower) band Green function. The chemical potential $\mu$ is determined by the doped carrier density $n$, 
\begin{eqnarray}
n =\frac{2}{V}\sum_{k}[-\int_{-\infty}^{\infty}\frac{d\omega}{\pi}F(\omega){\rm TrIm}G(k,\omega+i0) -2]\label{mu}
\end{eqnarray}
where the front factor 2 comes from the spin degree, the first term in the square brackets is the total occupation of electrons, the last term corresponds to the nondoped case with 2 as the valley degeneracy, and $F(\omega)$ is the Fermi distribution function. We hereafter will call $\mu$ determined by Eq. (\ref{mu}) as the renormalized chemical potential, distinguishing from the approximation $\mu \approx E_F = vk_F$ used in some cases such as the semiclassical Boltzmann theory at zero temperature.

We now consider the thermal transport. The (particle) current $J_1$ and heat current $J_2$ operators are defined as 
\begin{eqnarray}
J_1 &=& v\sum_k \psi^{\dagger}_{k}\tau_z\vec\sigma \psi_{k}  \nonumber\\
J_2 &=& iv\sum_k \psi^{\dagger}_{k}\tau_z\vec\sigma \frac{\partial}{\partial t}\psi_{k}. \nonumber
\end{eqnarray}
They correspond respectively to the forces $X_1 = -T^{-1}\nabla (\mu-e\phi)$ and $X_2 = \nabla (1/T)$ with $T$ as the temperature and $\phi$ the external electric potential.\cite{Mahan} The heat operator $J_2$ defined here is equivalent to devising the heat vertex as velocity$\times$frequency as shown by Johnson and Mahan \cite{Jonson} for the independent electrons interacting with the impurities. According to the linear response theory, the thermoelectric power is given by
$S = -L^{21}/eTL^{11}$
where the linear response coefficients $L^{\mu\nu}$ are obtained from the correlation function $\Pi^{\mu\nu}(\Omega^+)$ by 
\begin{eqnarray}
L^{\mu\nu} = -T\lim_{\Omega\to 0}{\rm Im}\Pi^{\mu\nu}(\Omega+i0)/\Omega.  \nonumber
\end{eqnarray}
In the Matsubara notation, $\Pi^{\mu\nu}(i\Omega_m)$ reads
\begin{equation}
\Pi^{\mu\nu}(i\Omega_m) = -\frac{1}{V}\int_0^{\beta}d\tau e^{i\Omega_m\tau}\langle T_{\tau}J_{\nu x}(\tau)J_{\mu x}(0)\rangle. \nonumber
\end{equation}
with $\beta = 1/T$. The quantity $L^{11} = T\sigma/e^2$ is related to the electric conductivity $\sigma$ which we have obtained in our previous work.\cite{Yan1}

Within the SCBA to the single particles, the correlation function $\Pi^{\mu\nu}$ is determined with the ladder-type vertex corrections. A common vertex $v\Gamma_x(k,\omega_1,\omega_2)$ given as the diagrams in Fig. 1(b) can be factorized out. $\Gamma_x(k,\omega_1,\omega_2)$ is expanded as\cite{Yan1}
\begin{equation}
\Gamma_x(k,\omega_1,\omega_2)=\sum_{j=0}^3y_j(k,\omega_1,\omega_2)A^x_j(\hat k) \nonumber
\end{equation}
where $A^x_0(\hat k)=\tau_z\sigma_x$, $A^x_1(\hat k)=\sigma_x\vec\sigma\cdot\hat k$, $A^x_2(\hat k)=\vec\sigma\cdot\hat k\sigma_x$, $A^x_3(\hat k)=\tau_z\vec\sigma\cdot\hat k\sigma_x\vec\sigma\cdot\hat k$, and $y_j(k,\omega_1,\omega_2)$ are determined by four-coupled integral equations.\cite{Yan} In the following, since the heat current-current correlation function will be analyzed for the case of $\omega_1 = \omega -i0\equiv \omega^-$ and $\omega_2 = \omega +i0\equiv \omega^+$, we here need to write out the relevant equations for this case. For briefness, we denote $y_j(k,\omega^-,\omega^+)$ simply as $y_j(k,\omega)$. To write in a compact form, we define the 4-dimensional vector $Y^t = (y_0,y_1,y_2,y_3)$ (where the superscript t implies transpose), and the matrices,
\begin{eqnarray}
U(k,k') = n_iv^2_0(|k-k'|)
\begin{pmatrix}
  1 &0 &0 &0\\
	0 &\cos\theta &0 &0\\
  0 &0 &\cos\theta &0\\
	0 &0 &0 &\cos 2\theta\\	
\end{pmatrix}\nonumber
\end{eqnarray}
where $\theta$ is the angle between $k$ and $k'$, and
\begin{equation}
M(k,\omega) = 
\begin{pmatrix}
  \bar g_{0}g_{0} &\bar g_{0}g_{c} &\bar g_{c}g_{0} &\bar g_{c}g_{c}\\
	\bar g_{0}g_{c} &\bar g_{0}g_{0} &\bar g_{c}g_{c} &\bar g_{c}g_{0}\\
  \bar g_{c}g_{0} &\bar g_{c}g_{c} &\bar g_{0}g_{0} &\bar g_{0}g_{c}\\
	\bar g_{c}g_{c} &\bar g_{c}g_{0} &\bar g_{0}g_{c} &\bar g_{0}g_{0}\\	
\end{pmatrix}\nonumber
\end{equation}
where $\bar g_{0,c}$ are complex conjugate of $g_{0,c} = g_{0,c}(k,\omega^+)$. The equation determining $Y(k,\omega)$ is then given by
\begin{equation}
Y(k,\omega) = Y_0 + \frac{1}{V}\sum_{k'}U(k,k')M(k',\omega)Y(k',\omega) \label{y}
\end{equation}
with $Y_0^t = (1,0,0,0)$.

We need to calculate the heat current-current correlation function $\Pi^{21}$ and then the coefficient $L^{21}$. The correlation function $\Pi^{21}$ is diagrammatically given by Fig. 1(c), which is a conserving approximation. From Fig. 1(c), we have
\begin{eqnarray}
\Pi^{21}(\Omega_m) &=& \frac{2v^2T}{V}\sum_{kn}i\omega_n{\rm Tr}[\tau_z\sigma_xG(k,i\omega_n)\nonumber\\
&&\Gamma_x(k,i\omega_n,i\omega+i\Omega_m)G(k,i\omega+i\Omega_m)]
\label{pin21}
\end{eqnarray}
where the factor 2 is due to the spin degeneracy, $i\omega_n$ (the fermionic Matsubara frequency) in front of the trace Tr operation comes from the heat vertex. The difference between $\Pi^{21}(\Omega_m)$ and $\Pi^{11}(\Omega_m)$ is there is the factor $i\omega_n$ in the above expression. According to the standard procedure,\cite{Mahan} by performing the analytical continuation $i\Omega_m \to \Omega + i0$ and taking the limit $\Omega \to 0$, one then gets a formula for $L^{21}$ in terms of the integral with respect to the real frequency, 
\begin{eqnarray}
L^{21} &=& \frac{T}{e^2}\int_{-\infty}^{\infty}d\omega[-\frac{\partial F(\omega)}{\partial\omega}]\omega\sigma(\omega)
\label{L21}
\end{eqnarray}
where $\sigma(\omega) = [P(\omega^-,\omega^+)-{\rm Re}P(\omega^+,\omega^+)]/2\pi$. Using the Ward identity, one obtains {\rm Re}$P(\omega^+,\omega^+) = -2e^2/\pi\hbar$,\cite{Yan} which contributes a constant term in $\sigma(\omega)$. The function $P(\omega^-,\omega^+)$ is given by
\begin{eqnarray}
P(\omega^-,\omega^+) &=& \frac{8v^2e^2}{V}\sum_{kj}M_{0j}(k,\omega)y_j(k,\omega),
\label{pr12}
\end{eqnarray}
with $M_{0j}(k,\omega)$ as the elements of the matrix $M(k,\omega)$ defined above. At low temperature, since the contribution to the integral in Eq. (\ref{L21}) comes from a small region around $\omega = 0$, one then expands $\sigma(\omega)$ as $\sigma(\omega) \approx \sigma(0)+\sigma'\omega$ with
\begin{eqnarray}
\sigma' &=& \frac{4v^2e^2}{\pi V}\sum_{kj}[\frac{\partial}{\partial\omega}M_{0j}(k,\omega)y_j(k,0)\nonumber\\
& & +M_{0j}(k,0)\frac{\partial}{\partial\omega}y_j(k,\omega)]_{\omega=0},
\end{eqnarray}
and obtains $L^{21} = T^3\pi^2\sigma'/3e^2$. The expression for the thermoelectric power then reads
\begin{eqnarray}
S = -\frac{\pi^2}{3}\frac{T\sigma'}{e\sigma} \label{s1}
\end{eqnarray}
in formal the same as the Mott relation.\cite{Mott} Since the quantity $\sigma'$ is involved with the functions $\partial g_{0,c}(k,\omega)/\partial\omega |_{\omega=0}$ and $\partial y_j(k,\omega)/\partial\omega |_{\omega=0}$, to obtain $S$ one needs to solve not only Eqs. (\ref{b1})-(\ref{mu}) and Eq. (\ref{y}) for functions $g_{0,c}(k,\omega)$ and $y_j(k,\omega)$ but also the equations for their frequency derivative at $\omega = 0$. The equations for $\partial g_{0,c}(k,\omega)/\partial\omega$ and $\partial y_j(k,\omega)/\partial\omega$ are given by the frequency derivative of Eqs. (\ref{b1})-(\ref{b4}) and Eq. (\ref{y}). 

\section{Numerical result}

\begin{figure} 
\centerline{\epsfig{file=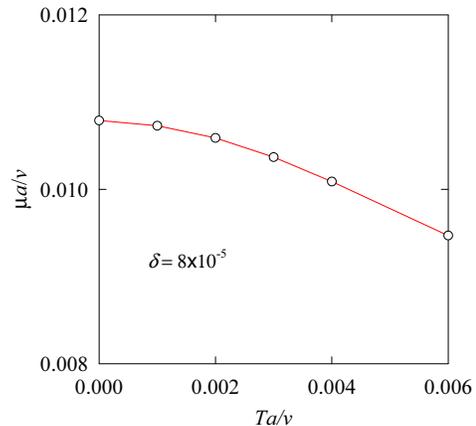,width=6.5 cm}}
\caption{(color online) Chemical potential $\mu$ as function of temperature $T$ at $\delta = 8\times 10^{-5}$.} 
\end{figure} 

In principle, the Green's function can be solved from Eqs. (\ref{b1})-(\ref{mu}) by iterations. However, there is difficulty in obtaining a convergent solution since in the intermediate iteration processes the Green's function is in usual not smooth and it is hard to satisfy Eq. (\ref{mu}). To overcome this difficulty, we perform the calculation of the Green's function at the Matsubara frequencies along the imaginary frequency axis for $T > 0$ using the method developed in Ref. \onlinecite{Yan4}. For doing so, Eqs. (\ref{b1})-(\ref{mu}) should be expressed for the Matsubara frequency. For example, Eq. (\ref{mu}) is rewritten as
\begin{eqnarray}
n &=&\frac{2}{V}\sum_{k}[T\sum_{\ell}{\rm Tr}G(k,i\omega_{\ell})e^{i\omega_{\ell}\eta}-2]\nonumber\\
  &=&\frac{2}{V}\sum_{k}\{T\sum_{\ell}{\rm Tr}[G(k,i\omega_{\ell})-G_0(k,i\omega_{\ell})]\nonumber\\
  & &-2[1-F(vk-\mu)-F(-vk-\mu)]\}
\label{mu1}
\end{eqnarray}
where $\omega_{\ell}$ is the fermionic Matsubara frequency, $\eta$ is an infinitesimal small positive quantity, and $G_0$ is the 0th Green's function. In the last equality, we have adopted the usual trick to improve the convergence of the series summation. In the present calculation, the parameters for sampling the Matsubara frequencies are $[h, L, M] = [2,20,5]$.\cite{Yan4} The number of the total frequencies is $L(M-1)+1 = 81$. The iteration is stable and converges fast. For zero temperature, $\mu$ can be determined by interpolation from the results at $T > 0$. As an example, we show in Fig. 2 the result for the chemical potential $\mu$ at the doped electron concentration $\delta = 8\times 10^{-5}$ (here $\delta$ is defined as the doped carriers per carbon atom $\delta = \sqrt{3}a^2n/4$). Usually, $\mu$ is a smooth function of the temperature $T$. With the chemical potential $\mu$ so obtained, the Green's function at real frequencies can be obtained by solving Eqs. (\ref{b1})-(\ref{b4}). 

We have numerically solved the integral equations for determining the functions $g_{0,c}(k,\omega)$, $y_j(k,\omega)$, $\partial g_{0,c}(k,\omega)/\partial\omega$ and $\partial y_j(k,\omega)/\partial\omega$ at $\omega = 0$ for various carrier concentrations. In Figs. 3-6, we show the self-energy $\Sigma_{\pm}(k,i0) = \Sigma_0(k,i0)\pm \Sigma_c(k,i0)$ (Fig. 3), the function $y_j(k,0)$ (Fig. 5) and their frequency derivative $\partial\Sigma_{\pm}(k,\omega+i0)/\partial\omega |_{\omega=0}\equiv \Sigma'_{\pm}(k,i0)$ (Fig. 4) and $\partial y_j(k,\omega+i0)/\partial\omega |_{\omega=0}\equiv y'_j(k,i0)$ (Fig. 6) for $\delta = 8\times 10^{-5}$ and $T = 0$. Notice that $\Sigma_{\pm}(k,\omega)$ correspond to the upper and lower band self-energies, respectively. The real part of the self-energy means the shift of the energy of the single particle, while the imaginary part is related to the lifetime. As seen from Fig. 3, overall, the upper band shifts downward but the lower band shifts upward. This change stems from the mixing of the states of two bands under the impurity scatterings. The functions $y_j$ reveal how the current vertex is modified by the impurity scatterings. For the bare current vertex only $y_0 = 1$ is finite. From Fig. 5, it is seen that the current vertex is significantly renormalized from the bare one. The functions $\partial\Sigma_{\pm}(k,\omega+i0)/\partial\omega |_{\omega=0}$ (Fig. 4) and $\partial y_j(k,\omega+i0)/\partial\omega |_{\omega=0}$ are structured around the Fermi wavenumber $k_F$. At larger $k$, they are smooth function of $k$.  

\begin{figure} 
\centerline{\epsfig{file=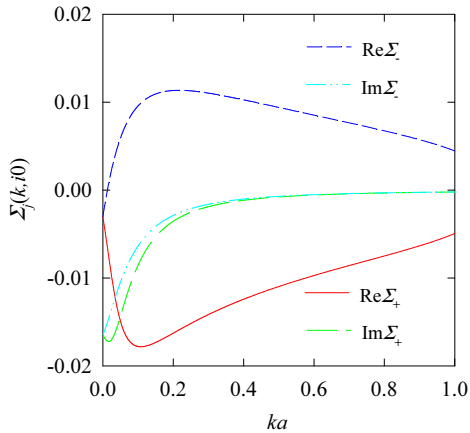,width=6.5 cm}}
\caption{(color online) Self-energy $\Sigma_{\pm}(k,i0)$ in unit of $v/a$ as function of $k$ at $\delta = 8\times 10^{-5}$ and $T = 0$.} 
\end{figure} 

\begin{figure} 
\centerline{\epsfig{file=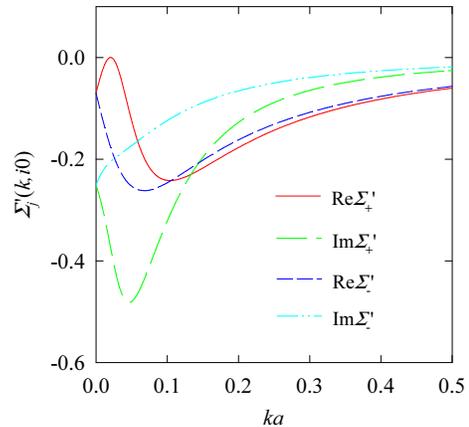,width=6.5 cm}}
\caption{(color online) Function $\partial\Sigma_{\pm}(k,\omega+i0)/\partial\omega |_{\omega=0}$ at $\delta = 8\times 10^{-5}$ and $T = 0$.} 
\end{figure} 

With the above results, the quantity $\sigma'$ and the thermoelectric power $S$ can be calculated accordingly. Shown in Fig. 7 are the obtained results for the quantity $\sigma'$ as function of the carrier concentration $\delta$. The circles are the fully self-consistent calculations with the chemical potential $\mu$ renormalized. For comparison, the result (squares) by the approximation $\mu \approx E_F$ is also plotted. $\sigma'$ increases with $\delta$ monotonically and is odd with respect to $\delta$ (electron) $\to$ -$\delta$ (hole). A notable feature is that $\sigma'$ varies dramatically within a narrow region $-\delta_0 \leq \delta \leq \delta_0$ with $\delta_0 = 7\times 10^{-5}$. Out of this region, the magnitude of $\sigma'$ increases with a slower rate as $|\delta|$ increasing. The inset in Fig. 7 shows the electric conductivity at low carrier concentration. The purple circles and the green squares are the interpolations. The values of the minimum conductivity so determined are 2.7 (in unit of $e^2/h$) for the renormalized $\mu$ and 3.5 for $\mu \approx E_F$, both of them larger than the well-known analytical result $4/\pi$ obtained from the single bubble using the phenomenological scattering rate in the Green's function.\cite{Shon,Ostrovsky} In a wide range of $\delta$, the overall behaviors of both results obtained using the renormalized $\mu$ and $\mu \approx E_F$ for the electric conductivity as function of $\delta$ are almost the same.\cite{Yan1} The dot-dashed line represents the extrapolation of $\sigma$ (for $\mu \approx E_F$) from large $\delta$. By carefully looking at the behavior of $\sigma$, we find that $\sigma$ starts to departure from the linearity at approximately the same $\delta_0$ below which $\sigma$ decreases slower as $\delta$ decreasing. 

\begin{figure} 
\centerline{\epsfig{file=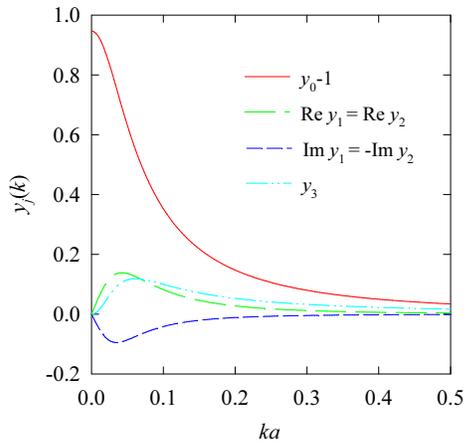,width=6.5 cm}}
\caption{(color online) Function $y_j(k,0)$ at $\delta = 8\times 10^{-5}$ and $T = 0$.} 
\end{figure} 

\begin{figure} 
\centerline{\epsfig{file=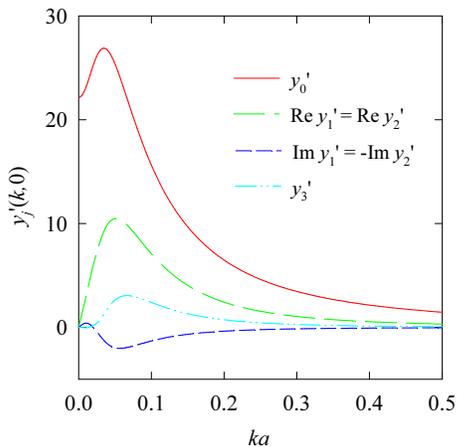,width=6.5 cm}}
\caption{(color online) Function $\partial y_j(k,\omega+i0)/\partial\omega |_{\omega=0}$ at $\delta = 8\times 10^{-5}$ and $T = 0$.} 
\end{figure} 
 
At low temperature, both $\sigma$ and $\sigma'$ are independent of $T$. Therefore, $S$ is a linear function of $T$ at low $T$. Shown in Fig. 8 are the numerical results for $S/T$ (red solid line with circles for the renormalized $\mu$ and the blue dashed line with squares for $\mu \approx E_F$) as function of $\delta$ and the comparison with the experimental measurements (symbols) by three groups.\cite{Zuev,Wei,Ong} Within the same narrow region $-\delta_0 \leq \delta \leq \delta_0$, the calculated $S/T$ varies drastically from the maximum at $\delta \approx -\delta_0$ to the minimum at $\delta \approx \delta_0$. Out of this region, the magnitude of $S/T$ decreases monotonically with $\delta$. Again, $S/T$ is an odd function of $\delta$. Clearly, the present calculation can capture the main feature of the experimental data. For the magnitude of $S/T$, there are obvious differences between the experimental results. This may be caused by the impurity distributions in samples treated by different experiments. 

\begin{figure} 
\centerline{\epsfig{file=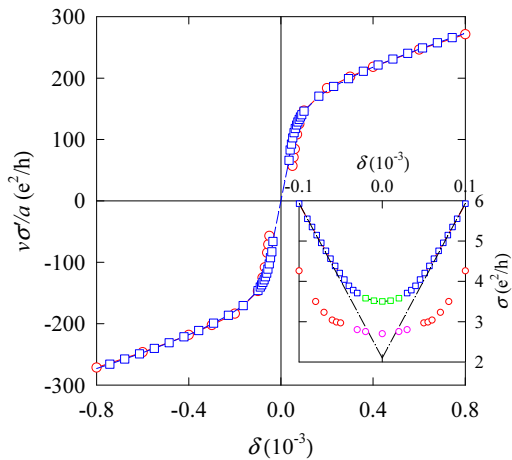,width=7. cm}}
\caption{(color online) Zero frequency derivative of the current-current correlation function as function of the carrier concentration $\delta$. The red circles and the blue squares are obtained by the calculations with the renormalized chemical potential $\mu$ and the one of $\mu \approx E_F$, respectively. Inset: The electric conductivity $\sigma$ at low carrier concentration. The purple circles and the green squares are the interpolations and the dot-dashed line is the extrapolation of the linear $\sigma$ (with $\mu \approx E_F$) in large $\delta$.} 
\end{figure} 

The features of $\sigma'$ and $S$ may be qualitatively explained by analyzing the behavior of $\sigma$ as function of $\delta$. Recall $\sigma = P(\omega^-,\omega^+)/2\pi|_{\omega = 0}+2e^2/\pi h$.\cite{Yan} $P$ is actually the functional of the Green function $G$ and the impurity potential $v_0$, both of latter two depending on $\delta$ or the chemical potential $\mu$. If the $\delta$ dependence of $v_0$ is neglected, then one gets $\sigma'$ from $d\sigma/d\mu$. Based on such a consideration, it has been illustrated in Ref. \onlinecite{Zuev} that the calculated $S$ from the experimental results for $\sigma$ is in overall agreement with experiment. Theoretically, at large carrier concentration, the system can be approximately described by the one band Green function, e.g., $g_+$ for electron doping. This is equivalent to the Boltzmann treatment to $\sigma$. On the other hand, the Boltzmann theory gives rise to a linear behavior of $\sigma$ down to very low $\delta$ close to 0. By the present formalism, however, there exists coherence between the states of upper and lower bands\cite{Trushin} at very low doping because the single particle energy levels are broadened under the impurity scatterings. The coherence is taken into account through the Green functions $g_{\pm}$ in the present formalism. At $\delta = \delta_0$, the coherence may be considered as setting in. As $\delta$ further decreases, the Fermi level gets close to the lower band and the coherence effect becomes significant. As a result, there is the minimum electric conductivity at $\delta = 0$. As seen from the inset in Fig .6, $\sigma$ decreases slower as $\delta < \delta_0$ being closer to zero, resulting in the rapid decreasing of $\sigma'$. The unusual behavior of $S$ at low doping comes from the combination of $\sigma$ and $\sigma'$ and can be understood as the coherence effect between the upper and lower Dirac bands.
 
\begin{figure} 
\centerline{\epsfig{file=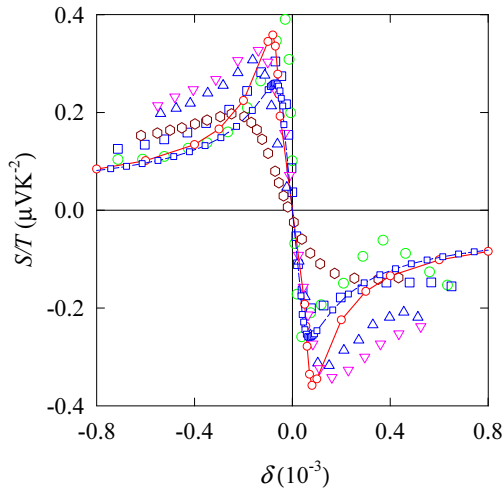,width=7. cm}}
\caption{(color online) Linear-$T$ dependence coefficient $S/T$ of thermoelectric power as function of the electron doping concentration $\delta$. The present calculations (red solid line with circles for renormalized $\mu$ and the blue dashed line with squares for $\mu \approx E_F$) are compared with the experimental data in Ref. \onlinecite{Zuev} (circles at $T = 150$ K and squares at $T = 300$ K), Ref. \onlinecite{Wei} (hexagons at $T = 255$ K), and Ref. \onlinecite{Ong} (up triangles at $T = 160$ K and down triangles at $T = 280$ K). } 
\end{figure}

The present model can not be applied to doping close to zero. At $\delta = 0$, there is no screening to the charged impurities by the model. This is unphysical. In a real system, there must exist extra opposite charges screening the charged impurities. This screening can be neglected only when above certain doping level the screening length by the carriers is shorter than that of the extra charges. Close to $\delta = 0$, the extra screening could be taken into account in a more satisfactory model. By the present model, we cannot perform numerical calculation at $\delta = 0$ because of the Coulomb divergence of $v_0(q)$ at $q = 0$. The minimum electric conductivity is obtained by interpolation. 

The unusual behavior of the thermoelectric power of graphene has also been studied recently by the semiclassical approach.\cite{Hwang1} For explaining the experimental observed transport properties of graphene at very low doping, Hwang {\it et al.} have proposed the electron-hole-puddle model.\cite{Hwang} By this model, the local carrier density is finite and the total transport coefficients are given by the averages of the semiclassical Boltzmann results in the puddles. The unusual behavior of $S$ and the minimum electric conductivity are so explained by the electron-hole-puddle model.  

At very low carrier doping, graphene is an inhomogeneous system as observed by experiment.\cite{Martin} There are regions where the carrier concentrations are very low. The resistance comes predominately from these regions. Our calculation at very low carrier concentration corresponds to studying the electron transport in these regions.

\section{Summary}

In summary, on the basis of self-consistent Born approximation, we have studied the thermoelectric power of Dirac fermions in graphene under the charged impurity scatterings. The current correlation functions are obtained by conserving approximation. The Green function and the current vertex correction, and their frequency derivative are determined by a number of coupled integral equations. The low-doping unusual behavior of the thermoelectric power at low temperature observed by the experiments is explained in terms of the coherence between the upper and lower Dirac bands. The present calculation for the thermoelectric power as well as for the electric conductivity is in very good agreement with the experimental measurements. 

\acknowledgments

This work was supported by a grant from the Robert A. Welch Foundation under No. E-1146, the TCSUH, the National Basic Research 973 Program of China under grant No. 2005CB623602, and NSFC under grant No. 10774171 and No. 10834011.

\end{document}